\documentclass[12pt, conference, onecolumn]{IEEEtran}
\IEEEoverridecommandlockouts
\usepackage{cite}
\usepackage{amsmath,amssymb,amsfonts}
\usepackage{graphicx}
\usepackage{textcomp}
\usepackage{xcolor}
\usepackage{url}
\definecolor{gray}{rgb}{0.95,0.95,0.95}
\definecolor{darkgray}{rgb}{0.4,0.4,0.4}
\definecolor{lgray}{RGB}{250,250,250}
\definecolor{lgreen}{RGB}{63,127,95}
\definecolor{lred}{RGB}{127,0,85}
\definecolor{lblue}{RGB}{42,0,255}
\definecolor{orange}{rgb}{1,0.5,0}

\usepackage{listings} 
\def\BibTeX{{\rm B\kern-.05em{\sc i\kern-.025em b}\kern-.08em
    T\kern-.1667em\lower.7ex\hbox{E}\kern-.125emX}}

\usepackage{tikz,xcolor,hyperref}
\definecolor{lime}{HTML}{A6CE39}
\DeclareRobustCommand{\orcidicon}{%
	\begin{tikzpicture}
	\draw[lime, fill=lime] (0,0) 
	circle [radius=0.16] 
	node[white] {{\fontfamily{qag}\selectfont \tiny ID}};
	\draw[white, fill=white] (-0.0625,0.095) 
	circle [radius=0.007];
	\end{tikzpicture}
	\hspace{-2mm}
}

\foreach \x in {A, ..., Z}{%
	\expandafter\xdef\csname 
	orcid\x\endcsname{\noexpand\href{https://orcid.org/\csname 
	orcidauthor\x\endcsname}{\noexpand\orcidicon}}
}

\begin{document}
	\bibliographystyle{IEEEtran}

\title{Simple Spyware\\ Androids Invisible Foreground Services and How to (Ab)use 
	Them}

\author{
\IEEEauthorblockN{Whitepaper \\ Thomas Sutter\orcidA{}}
\IEEEauthorblockA{
Winterthur, Switzerland \\
suth@zhaw.ch\\
December 2019
}
}
\maketitle

\begin{abstract}
With the releases of Android Oreo \cite{AndroidOreoBackgroundLimitations} and 
Pie \cite{AndroidPieReleaseNotes}, Android introduced some background 
execution limitations for apps. Google restricted the execution of background 
services to save energy and to prevent apps from running endlessly in the 
background. Moreover, access to the device's sensors was changed and a new 
concept named foreground service has been introduced. Apps were no longer 
allowed to run background services in an idle state, preventing apps from using 
the device's resources like the camera. These limitations, however, would not 
affect so-called foreground services because they show a permanently visible 
notification to the user and could therefore be stopped by the user at any time. Our 
research found out that flaws in the API exists, which allows starting invisible 
foreground services, making the introduced limitations ineffective. We will show 
that the found flaws allow attackers to use foreground services as a tool for spying 
on users.
\end{abstract}

\section{Introduction}
We found out that foreground services do not show any visual notification when 
the service's execution time is shorter than five seconds\footnote{The exact 
duration is depending on the phone.}. We use this loading time and combine it with 
another flaw in Androids Job Scheduler API to continuously execute tasks from a 
background context.
Exploiting these flaws allows apps to use the device's resources, even when the 
app is closed or on standby. Furthermore, we show that we can use these flaws 
for continually spying on users and allowing malware developers to create spyware 
without the need for complicated exploitation.

We start in Section \ref{abusing} with some basic introduction to Android's 
components and then we explain how we spawn a foreground service 
on Android Pie combined with some basic background schedulers. We then 
use this basic example to show how we can use these API's to implement a simple
spyware app. At the end of the paper in Section \ref{results} we discuss some 
limitations as well as ideas to prevent such attacks.

\section{Abusing Foreground Services} \label{abusing}
Android defines two basic types of service classes for apps, background and 
foreground services \cite{ServicesOverview}. The difference between these 
services is how they appear in the user interface and under which constraints they 
are executed. We can start Foreground services even when the user interface is 
closed; in contrast, background services cannot do so. The operating system 
prevents background services to start when the app's user interface is closed by 
throwing an \textit{IllegalStateExeception}. If we want to run a task when the app 
is closed, we can use Android's scheduling classes instead. For example, the 
JobScheduler \cite{JobSchedulerClass} and AlarmManager 
\cite{AlarmManagerClass} classes.

Those schedulers' idea is that apps can synchronize or process data even when 
the user interface gets closed. We can use this, for example, to set an alarm 
clock or to upload a file when the user interface is not needed. If misused, 
schedulers often use a lot of battery power. For example, when an app is 
continuously uploading data in the background. Since Android Oreo and app can 
no longer run endless background services when 
its user interface is not shown. Usually, the operating system stops all services 
some minutes after the app was closed. Moreover, access to sensors like 
microphones and cameras should no longer work when the app is closed. In case 
an app tries to access one of the restricted sensors from a scheduler, the 
operating system throws an \textit{IllegalStateException}, and the access is 
thereby not 
granted. However, in order to access the sensors from the background context, 
foreground services can be used. Schedulers are allowed to start foreground 
services, and as mentioned, foreground services do not have any restrictions 
when it comes to sensor access. The only limitations 
foreground services have is that they need to show a permanently visible 
notification and that the app spawning the foreground service needs to have the 
permission to access the sensor. Listing \ref{ForegroundServiceCode} shows an 
example code to start a foreground service in Java.

\begin{lstlisting}[label=ForegroundServiceCode, frame=single, language=Java, 
caption=Java code to spawn and abuse foreground service.]  
public class SomeExampleService extends Service {
	// Start service by intent. No filtering shown here
	@Override
	public int onStartCommand(Intent intent, int flags, int startId){
		// ~4.9999.. seconds to call startForeground(...)
		Notification notification = createCustomNotification();
		this.startForeground(1, notification) // Sensor access not restricted anymore 
		// Race condtion started. Let's collect some data fast...
		accessCamera();
		accessMicrophone();
		// ... some more malicious code
		stopForeground(true); //Stop the service before notification is loaded and 
		// win the race against the notification manager.
		return START_STICKY;
	}
...
	/**
	* Create a custom notification.
	*/
	private Notification createCustomNotification() {
		NotificationChannel channel = new NotificationChannel("1", "Location", 
		NotificationManager.IMPORTANCE_LOW);
		NotificationManager notificationManager = getSystemService(NotificationManager.class);
		if(notificationManager != null){
			notificationManager.createNotificationChannel(channel);
			NotificationCompat.Builder builder = new NotificationCompat.Builder(this, "1")
			.setSmallIcon(R.drawable.ic_remove_red_eye_black_24dp)
			.setContentTitle("Simple Spyware")
			.setContentText("Tracking your position!")
			.setPriority(NotificationCompat.PRIORITY_LOW)
			return builder.build();
		}
	}
}
\end{lstlisting}

\begin{figure}[h]
	\includegraphics[width=\linewidth]{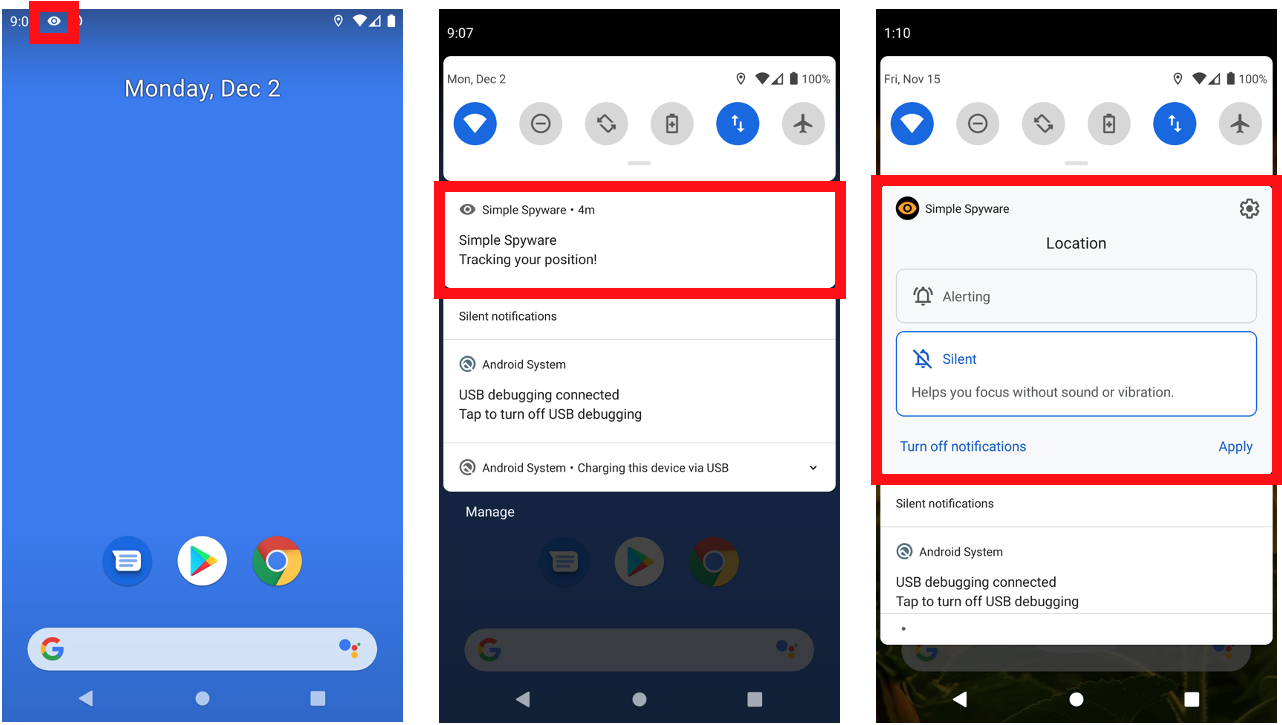}
	\caption{Example of foreground notifications in the user interface.}
	\label{fig:01_Notifications}
\end{figure}

What we can see in Listing \ref{ForegroundServiceCode} is that we extend the 
service class and that we define a \textit{onStartCommand} method in Line 6 as 
we would do with a regular foreground service. To start a foreground service, we 
have to define a notification. On Lines 20 to 31, we define the notification and set 
an example icon and text. As shown, we can customize the appearance of the 
notification as we would like it to be. Figure \ref{fig:01_Notifications} shows how 
such a foreground service in the user interface could look. 

The notification is \textit{sticky}, and we cannot dismiss it until we stop foreground 
service or the user manually disables the notification. Attempting to cancel the 
notification within Java will not work since the notification manager will not allow it. 
An attacker could hide its intention by showing some typical notification like an 
update, announcement, or loading screen. In many cases, as long as the 
notification does not stay too long in the notification bar, it will not raise the user's 
suspicion. Nevertheless, users could get suspicious or annoyed when the 
notification shows them some unwanted content or the notification stays too long 
in the notification bar. In such cases, the user can stop the app or disable the 
notifications for the app. 

As attackers, we wanted to go a step further and see if we could stop the 
notification from showing up at all, so that the users would not get alarmed. So we 
search for 
possibilities to dismiss or cancel the notification but could not find an easy way to 
do it. Instead, we found out that the notification manager does not show a 
notification when the foreground service's lifetime is shorter than five seconds. We 
can use this race condition to execute any code before the operating system loads 
the notification.

We can spawn one invisible foreground service and execute our code once, 
and then the app will stop. To further abuse this approach, we needed a way to 
persistently spawn new foreground services on demand. To do so we use 
Android's JobScheduler \cite{JobSchedulerClass} or AlarmManager 
\cite{AlarmManagerClass} classes which allows to execute code outside of the 
app context. 
For decided to use for our example the JobScheduler class but the 
AlarmManager has more or less the same functionality. We create a new Job with 
the JobInfo.Builder \cite{AndroidJobSchedulerBuilder} as shown in Listing 
\ref{JobScheduler}.

\newpage
\begin{lstlisting}[frame=single, language=Java, label=JobScheduler, caption=Java 
example for the job scheduler class.]  
public void scheduleJob(){
    long interval = 1000 * 60L;	// Some interval
    ComponentName serviceComponent = new ComponentName(this, 
    JobScheduler.class);
    JobInfo.Builder builder = new JobInfo.Builder(JOB_ID, serviceComponent);
    builder.setPeriodic(interval);	// Minimum is 15 minutes
    builder.setOverrideDeadline(interval * 2);	// Sets the maximum scheduling 
    // latency
    builder.setMinimumLatency(interval);	// Runs a job after a delay
    JobScheduler jobScheduler = this.getSystemService(JobScheduler.class);
    jobScheduler.schedule(builder.build());	// Schedule the job
}
\end{lstlisting}

As attackers, we want to use the JobScheduler to start a job every X seconds or 
minutes. Our idea is that we want to execute any malicious command like taking a 
picture or uploading a file whenever we need it. The JobScheduler class has a 
method \textit{.setPeriodc(long seconds)} that offers precisely that. The problem 
with this method is that it has a minimum interval of 15 minutes. So if we would 
use it, we could only execute malicious code every 15 minutes. For some 
malicious commands, this is maybe a too long period, so we wanted to circumvent 
this limitation. Instead of using the .setPeriodic method, we can create a custom 
scheduler. All we have to do so is to use the JobScheduler's 
\textit{.setMinimumLatency} method. This method allows us to run a job after a 
given delay and has no limitations in execution time. In other words, we can set a 
delayed job under 15 minutes an circumvent the limitation.
Consequently, we can build our period job scheduler by chaining jobs with the 
delay function. Whenever we execute a job, we schedule a new job directly with 
the delay method. As long as our job chain is not interrupted, we can use the 
JobScheduler class to spawn new foreground services at demand. The chaining of 
jobs in combination with foreground services allows us to circumvent the 
background service limitation introduced in Android Oreo 
\cite{AndroidOreoBackgroundLimitations} and the background sensor access 
limitation introduced in Android Pie \cite{AndroidPieReleaseNotes}. 

If we want, we can further enhance our scheduling with some other options of the 
JobScheduler class. For example, we could only schedule jobs when the device is 
charging or connected to wifi. We can as well change our scheduling strategy 
during execution if necessary, to stay undetected.

\begin{itemize}
	\item setPersisted(boolean isPersisted): This allows that a job persists restarts. 
	Needs the \textit{received\_boot\_completed} permission to do so.
	
	\item setRequiredNetwork(NetworkRequest networkRequest) and 
	setRequiredNetworkType(int networkType): This allows us to define a specific 
	network type to be active before the job is executed.
	
	\item setRequiresBatteryNotLow(boolean batteryNotLow): Run a job only if the 
	battery is not low. Usually, this is when the phone has more than 15\% capacity.
	
	\item setRequiresCharging(boolean requiresCharging): Run a job only when the 
	device is charging.
	
	\item setRequiresDeviceIdle(boolean requiresDeviceIdle): Run a job only if a 
	device is not used and therefore in idle state.
\end{itemize}
See \cite{JobSchedulerClass} for a complete overview of methods. An example of 
this approach was implemented in our open-source demo app  
\cite{SimpleSpywareGit}. This approach also works with the AlarmManager class, 
and the code is as well available in our git.

\section{Collecting Data} 
After we have set up the chain of jobs and start our foreground service, we can 
add methods to collect the user's data. As shown in Listing 
\ref{ForegroundServiceCode} on Line 9, it is common for spyware to take camera 
pictures or to record the microphone audio. We have tested if our approach works 
with these features, and we implemented some examples in our demo app.

\subsection{Camera} \label{cameraAbuse}
We can modify some of the existing open-source libraries \cite{HiddenCameraGit, 
AndroidSpyCamera, AndroidSpyCamera2} for taking hidden camera pictures to 
demonstrate that we can access the camera API from the background. We start 
an invisible foreground service and use our access to the camera2 API to capture 
some images. In most cases and with the phones we had at hand, this approach 
works as expected. However, during testing, we noticed that our hidden camera 
implementation does not work on all cameras. Some cameras only show black 
images due to a short exposure time or wrong camera calibration.

\subsection{Microphone}
As we have explained, we use short lived foreground services with an execution 
time of maximal five seconds. This execution time is for most audio recordings 
not sufficient, and we would generate most likely incomplete audio recordings if we 
tried to record only for five seconds. However, it is possible to record audio for a 
long time due to Android's Media Server architecture \cite{MediaAPI}. Usually, if 
we want to record audio, we use the MediaRecorder class 
\cite{MediaRecorderAPI}, and we can start to record with the \textit{.start()} 
method. As soon as we do, our app will contact Android's media server, and the 
server will start recording. Since our app's background limitations do not apply to 
the media server process, it can record the audio even if we close our spyware 
app. Consequently, our app needs only to control when the audio recording needs 
to be started or stopped, and we can do this within the five seconds execution 
time of our invisible foreground service.

\subsection{Location}
Location tracking is another interesting feature since not only spyware developers 
are interested in collecting this data. Tracking the position with an invisible 
foreground service has the advantage that there is no limitation on how often we 
can get position updates from the location API. According to 
\cite{AndroidBackgroundLocationLimits} Background services are limited to 
request location updates a few time per hour. Consequently, when we use 
foreground services, we can track the location near real-time with high precision, 
which is interesting not only for spyware. However, If we track the position in 
real-time, it will consume a lot of battery power, and battery optimization is likely 
to trigger and alarm the user.

\begin{figure}[h]
	\centering
	\includegraphics[width=0.8\linewidth]{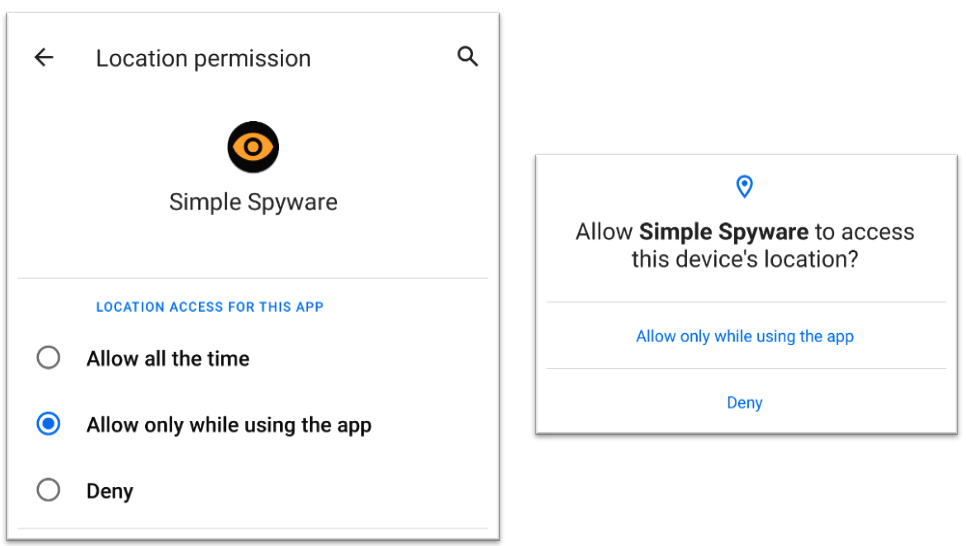}
	\caption{Android10 new runtime location permission.}
	\label{fig:02_Android10Location}
\end{figure}

When Android10 was released, we wanted to test if our approach still worked on 
the new Android version. Google introduced with Android10 a new permission level 
called "Allow only while in use" for location tracking \cite{Android10Release}. 
Moreover, since the release, new apps had to add a new permission to their 
manifest, \textit{ACCESS\_BACKGROUND\_LOCATION}, if they want to access 
the location from a background service. However, tracking the location within a 
foreground services does not need this permission as long as the service type 
within the manifest is set to \textit{location} (see Listing 
\ref{ManifestLocation}). 

\begin{lstlisting}[frame=single, language=Java, label=ManifestLocation, 
caption=Manifest 
example for the new foreground service type.]
<service android:name=".data.location.LocationTrackerService"
			   android:foregroundServiceType="location">
\end{lstlisting}
We added the new type in the manifest and tested it. First, we tested it with the 
"Allow all the time" permission, and as expected, we were able to track the 
location within our invisible foreground service. As a second test, we tested if the 
"Allow only while in use" permission would have any effect since we think Google 
created it to prevent what we were doing precisely. However, we could still track 
the location of the phone. The system sees a foreground service by definition to 
be in the foreground and, therefore, to be used.  As a consequence, we can use 
an invisible foreground service and track the location even when the app gets 
closed and the "Allow only while in use" permission is set.

\subsection{Files \& Others}\label{fileAbuse}
As the last attack, we wanted to see if we could use our foreground service to 
monitor the file storage. Similar to the other attacks, we can conduct this one 
by running a foreground service every X minutes. In our case we just tested if it is 
possible to upload files from the external storage like camera pictures without the 
notification showing up. We found out that it is possible for smaller files if the 
data-connection is fast enough. We could use the .setRequiresCharging and the 
.setRequiredNetwork method to further improve our uploading strategy and upload 
larger files when the phone has the resources to do so.

When an app does not have the permissions to access the camera or the location, 
it may have the file permission. Android Oreo and Pie allow an app to read all the 
content on the external storage where often private critical files like photos or 
documents are store. An attacker can then read the information stored in the 
external storage and use a foreground service to upload documents to a remote 
server.

We think this is primarily a problem when all pictures from the standard camera 
app are stored on the external storage by default. Attackers then get direct access 
to all user camera pictures, which is a privacy problem. Moreover, if the user has 
activated the GPS metadata tracking on the camera app, all pictures will have the 
location information. So if we have the file permission, an attacker can get access 
to the camera pictures and where the user has taken the pictures. Anti-virus 
vendors have reported such attacks in the past \cite{EXIFExample}.

We decided not to integrate this feature into our demonstration app because we 
would have needed to add internet permission to our app, and we want it to be 
safe for testing usages. Furthermore, we tested clip-board hijacking and overlay 
attacks, which also work with invisible foreground services, and decided not to 
integrate them into our demo app.

\section{Limitations}
In Sections \ref{cameraAbuse} to \ref{fileAbuse} we demonstrated just some 
examples of possible attacks. All of the common attacks are possible by using a 
combination of background and foreground services. Nevertheless, we want to 
discuss as well the limitations of our approach.

First of all, we tested our approach only on three Android 
devices\footnote{Samsung Galaxy S9+, Samsung A50, and Huawei P smart} and 
some Android Oreo, Pie, and Android10 emulators. Since wide-scale testing is not 
feasible within this project, the foreground services may behave differently on 
other phones. Some vendors may have additional security measurements 
implemented, which can defeat or detect invisible foreground services. As far as 
we know, it works on all tested Android Oreo and Pie phones and as well on 
Android10 devices.

Second, we know that the access to some sensors is restricted to one app at a 
time: For example, if the user has already occupied the camera, a foreground 
service cannot access the camera for spying. Depending on the persistence 
strategy, this can occur more often than one may think and can attract the user's 
attention. For example, if our spyware captures a picture and the user has FaceID 
activated, it can occur that FaceID cannot access the camera and show an error 
message.

Third, schedulers may not run at specific times: Android reschedule jobs and 
alarms, for example, when the device goes to the idle state. If our device is for a 
more extended period in stand-by, the schedulers will likely not execute our code 
until we use the device again. Furthermore, scheduling strategies can differ from 
device to device, and therefore the task execution can work correctly on one 
device but not on another.

Fourth, some vendors have different behavior for showing notifications. For 
example, the location icon on some phones is shown as soon as the user 
activates location tracking and is permanently visible in the top menu bar. Other 
vendors only show the location icon whenever an app accesses the phone's 
location. So visibility for some features is different on some devices.

Fifth, if we choose a spying strategy that uses many phone resources, like taking 
a camera picture every 10 seconds, it is likely that the operating system's battery 
optimizations will trigger. Depending on the phone vendor, it may show a 
notification to the user or directly stop the execution of our app.

Sixth, all the demonstrated malware features work only if the user has already 
installed the app and has given the app the necessary runtime permissions. 

\section{Results} \label{results}
Our demo spyware shows that Android's permission model cannot prevent 
excessive use of permissions and that the limitations do not prevent the collection 
of the user's sensitive data. As we described, the access restriction in Android Pie 
cannot entirely prevent our access to the critical sensors like the camera over 
foreground services. We think the restrictions, in general, are a good idea and give 
the user more security, but it still lacks some fundamental points like restricting 
the file access. We can summarize what we have done in the following steps:

\begin{enumerate}
	\item \textbf{Background Scheduling:} We frequently spawn new jobs with a 
	chain of 
	JobScheduler or AlarmManager jobs. With every job we execute, we start a new 
	short-lived foreground service.
	
	\item \textbf{Invisible Foreground Service:} As long as our foreground service's 
	execution 
	time is shorter than five seconds\footnote{Five seconds it just the average. 
	Timing may change on other phones.}, it will not display a notification. The 
	operating 
	system allows only foreground service to access critical sensors like cameras 
	and 
	microphones.
	
	\item \textbf{Spying:} We can use common spying techniques to collect data 
	from the 
	user during the five-second window. We tested taking camera pictures, 
	recording 
	audio, uploading files, or tracking the user's location and other features. We 
	have 
	shown that our approach works as well on Android10.
\end{enumerate}

Patching these issues is not as simple as it may seem. Since invisible foreground 
services only use standard API calls, which are unlikely to be removed soon. 
Therefore we think that such attacks are likely to be seen for a longer time. Even 
if a patch for the foreground service is released and the notification's behavior 
changes, it seems that an attacker still has some possibilities for workarounds. 
Access and timings may get patched, but we think it will not entirely prevent such 
attacks since an attacker can still set the notification design. Attackers maybe will 
come up with custom notifications that do not look suspicious to the user.

\section{Discussion}

\subsection{Transparency}\label{monitoring}

If we wanted to prevent such attacks with the current design, we could 
continuously monitor the apps' permission usage. In case we would find an app 
that misused its permission, we could revoke the permissions. Permission 
monitoring apps exist but are often not accurate or widely used. For example, 
Samsung's "App Permission Monitor"\cite{SamsungAppMonitor} logs the access, 
and the app notifies the 
user as soon as the monitor detected suspicious permission access. Monitoring 
the permission usage of an app can help detect abusive apps, but monitoring is 
often error-prone, and automated detection is difficult. A malicious app may only 
use permissions sporadically and, therefore, stay undetected, or some apps need 
to access some permissions more often than others. In general, monitoring is 
unlikely to solve the problem entirely, but it could give transparency to what is 
running in the background to security-aware users.

Another problem we should address is that Android users cannot check which 
tasks are running in the background. Staying hidden from the user allows 
malicious apps to run code unseen. Even if the user does not grant any dangerous 
permission, it allows an attacker to collect the phone's usage data, such as the 
installed apps list.

\subsection{Revocation}
We can argue that users can deny dangerous permissions to prevent the 
described attacks in Section \ref{abusing}. However, the common problem is that 
many users do not fully understand which permissions are essential for an app to 
work and grant the requested runtime permissions. We think this leads to the fact 
that some apps have too many permissions that they do not need to function 
correctly. Another point is that once we give permission, the app holds access as 
long as it does not revoke it. In many cases, an app needs dangerous permission 
just a couple of times when the user uses the app. We suggest that the operating 
system automatically revokes permissions when they are not needed and that 
Android implements one time and time-based permissions. We think this could 
help in mitigating the time-frame a malicious app can collect data of the users. We 
can argue that one-time permissions or automatic permission revocation is not 
user-friendly in terms of usability, but not giving Android users the possibility to 
defend themselves against such fraud is it neither.

\section*{Acknowledgment}
I want to thank my professor Dr. Bernhard Tellenbach for encouraging me to 
publish this work. Moreover, I would like to thank the Zurich University of Applied 
Science for supporting my research on this topic.

\bibliography{iteratur}

\begin{thebibliography}{10}
\providecommand{\url}[1]{#1}
\csname url@samestyle\endcsname
\providecommand{\newblock}{\relax}
\providecommand{\bibinfo}[2]{#2}
\providecommand{\BIBentrySTDinterwordspacing}{\spaceskip=0pt\relax}
\providecommand{\BIBentryALTinterwordstretchfactor}{4}
\providecommand{\BIBentryALTinterwordspacing}{\spaceskip=\fontdimen2\font plus
\BIBentryALTinterwordstretchfactor\fontdimen3\font minus
  \fontdimen4\font\relax}
\providecommand{\BIBforeignlanguage}[2]{{%
\expandafter\ifx\csname l@#1\endcsname\relax
\typeout{** WARNING: IEEEtran.bst: No hyphenation pattern has been}%
\typeout{** loaded for the language `#1'. Using the pattern for}%
\typeout{** the default language instead.}%
\else
\language=\csname l@#1\endcsname
\fi
#2}}
\providecommand{\BIBdecl}{\relax}
\BIBdecl

\bibitem{AndroidOreoBackgroundLimitations}
\BIBentryALTinterwordspacing
Google. (2017, May) \BIBforeignlanguage{English}{Background execution limits}.
  Google Corporation. [Online]. Available:
  \url{https://developer.android.com/about/versions/oreo/background#services}
\BIBentrySTDinterwordspacing

\bibitem{AndroidPieReleaseNotes}
\BIBentryALTinterwordspacing
------. (2019, April) \BIBforeignlanguage{English}{Behavior changes: all apps}.
  Google Corporation. [Online]. Available:
  \url{https://developer.android.com/about/versions/pie/android-9.0-changes-all}
\BIBentrySTDinterwordspacing

\bibitem{ServicesOverview}
\BIBentryALTinterwordspacing
------, ``Services overview,'' January 2019. [Online]. Available:
  \url{https://developer.android.com/guide/components/services}
\BIBentrySTDinterwordspacing

\bibitem{JobSchedulerClass}
\BIBentryALTinterwordspacing
------. (2019, April) \BIBforeignlanguage{English}{Jobscheduler}. Google
  Corporation. [Online]. Available:
  \url{https://developer.android.com/reference/android/app/job/JobScheduler}
\BIBentrySTDinterwordspacing

\bibitem{AlarmManagerClass}
\BIBentryALTinterwordspacing
------. (2019, April) \BIBforeignlanguage{English}{Alarmmanager}. Google
  Corporation. [Online]. Available:
  \url{https://developer.android.com/reference/android/app/AlarmManager?hl=en}
\BIBentrySTDinterwordspacing

\bibitem{AndroidJobSchedulerBuilder}
\BIBentryALTinterwordspacing
------. (2019, April) \BIBforeignlanguage{English}{Jobinfo.builder}. Google
  Corporation. [Online]. Available:
  \url{https://developer.android.com/reference/android/app/job/JobInfo.Builder}
\BIBentrySTDinterwordspacing

\bibitem{SimpleSpywareGit}
\BIBentryALTinterwordspacing
T.~Sutter. (2019, December) \BIBforeignlanguage{English}{Simple spyware github
  code}. [Online]. Available:
  \url{https://github.com/7homasSutter/SimpleSpyware}
\BIBentrySTDinterwordspacing

\bibitem{HiddenCameraGit}
\BIBentryALTinterwordspacing
K.~Patel. (2016, April) \BIBforeignlanguage{English}{android-hidden-camera}.
  [Online]. Available:
  \url{https://github.com/hussainbadri21/android-hidden-camera}
\BIBentrySTDinterwordspacing

\bibitem{AndroidSpyCamera}
\BIBentryALTinterwordspacing
TwoEightNine. (2016, April)
  \BIBforeignlanguage{English}{android-hidden-camera}. [Online]. Available:
  \url{https://github.com/TwoEightNine/AndroidSpyCamera}
\BIBentrySTDinterwordspacing

\bibitem{AndroidSpyCamera2}
\BIBentryALTinterwordspacing
hzitoun. (2017, October)
  \BIBforeignlanguage{English}{android-camera2-secret-picture-taker}. [Online].
  Available:
  \url{https://github.com/botyourbusiness/android-camera2-secret-picture-taker}
\BIBentrySTDinterwordspacing

\bibitem{MediaAPI}
\BIBentryALTinterwordspacing
Google. (2019, October) \BIBforeignlanguage{English}{Media app architecture
  overview}. Google Corporation. [Online]. Available:
  \url{https://developer.android.com/guide/topics/media-apps/media-apps-overview}
\BIBentrySTDinterwordspacing

\bibitem{MediaRecorderAPI}
\BIBentryALTinterwordspacing
------. (2019, October) \BIBforeignlanguage{English}{Mediarecorder overview}.
  Google Corporation. [Online]. Available:
  \url{https://developer.android.com/guide/topics/media/mediarecorder}
\BIBentrySTDinterwordspacing

\bibitem{AndroidBackgroundLocationLimits}
\BIBentryALTinterwordspacing
------. (2019, October) \BIBforeignlanguage{English}{Privacy changes in android
  10}. Google Corporation. [Online]. Available:
  \url{https://developer.android.com/about/versions/oreo/background-location-limits}
\BIBentrySTDinterwordspacing

\bibitem{Android10Release}
\BIBentryALTinterwordspacing
------. (2019, October) \BIBforeignlanguage{English}{Privacy changes in android
  10}. Google Corporation. [Online]. Available:
  \url{https://developer.android.com/about/versions/10/privacy/changes}
\BIBentrySTDinterwordspacing

\bibitem{EXIFExample}
\BIBentryALTinterwordspacing
J.~Lister. (20.11.2019) Android malware records calls, tracks location.
  [Online]. Available:
  \url{https://www.infopackets.com/news/10657/android-malware-records-calls-tracks-location}
\BIBentrySTDinterwordspacing

\bibitem{SamsungAppMonitor}
\BIBentryALTinterwordspacing
Samsung. (2019, October) \BIBforeignlanguage{English}{What is app permission
  monitor feature and how to turn it off?} [Online]. Available:
  \url{https://www.samsung.com/ae/support/mobile-devices/what-is-this-new-add-feature-app-permission-monitor-and-how-to-turn-off/}
\BIBentrySTDinterwordspacing

\end{thebibliography}
\end{document}